# Transient Execution of Non-Canonical Accesses.

Saidgani Musaev
Christof Fetzer
TU Dresden
Germany

## ABSTRACT

**Recent years have brought microarchitectural security into the spotlight, proving that modern CPUs are vulnerable to several classes of microarchitectural attacks. These attacks bypass the basic isolation primitives provided by the CPUs: process isolation, memory permissions, access checks, and so on. Nevertheless, most of the research was focused on Intel CPUs, with only a few exceptions. As a result, few vulnerabilities have been found in other CPUs, leading to speculations about their immunity to certain types of microarchitectural attacks. In this paper, we provide a black-box analysis of one of these under-explored areas. Namely, we investigate the flaw of AMD CPUs which may lead to a transient execution hijacking attack. Contrary to nominal immunity, we discover that AMD Zen family CPUs exhibit transient execution patterns similar for Meltdown/MDS. Our analysis of exploitation possibilities shows that AMDs design decisions indeed limit the exploitability scope comparing to Intel CPUs, yet it may be possible to use them to amplify other microarchitectural attacks.**

## 1 INTRODUCTION

Modern processors implement Out-of-Order (OOO) and Speculative Execution to achieve better performance and higher instruction-level parallelism. These features often lead to issuing wrong instructions into the processor pipeline. The reasons for the wrong instruction stream may be a misprediction on speculation, an exception handling delay, or a microcode-assisted event. Although these instructions never change the architectural state, they may be still executed on the CPU pipeline. Such execution of "invisible" instructions is called transient execution. It is possible to detect the presence of such execution via microarchitectural state changes (memory caches). It has been repeatedly shown [1, 3, 4, 6, 8, 9, 11, 13, 15, 19, 22] that many microarchitectural features can trigger violations of software security boundaries during transient execution. These attacks are represented by two big classes. The first is *LVI and Spectre*-type attacks. This class targets specific instruction gadgets in the victim domain, by "poisoning" microarchitectural state of the processor to direct transient (speculative, out-of-order, microcode-assisted) execution path, which allows an attacker to control the victim's execution and to create a covert channel. The second is *Meltdown*-type attacks. This class targets architecturally illegal data flow from microarchitectural elements (e.g., L1 Cache, Store/Load-Buffer, Special Register Buffer). Such an illegal data flow allows an attacker to exploit transient execution to expose data and change the microarchitectural state. Although, *Meltdown*-type attacks are represented by the separate class, *Spectre*-type attacks may be built on nop of the same primitives as *Meltdown*-type attacks. While Spectre-type attack targets wide families of CPUs from different vendors, Meltdown-type attacks were targeting mostly Intel CPUs.

The predominant focus of previous research on Intel may mean that other vendors' CPUs were not investigated as thoroughly and may still have undiscovered microarchitectural vulnerabilities. Therefore, we investigated AMD CPUs as they have a similar architecture yet received less attention from the security community. Even though some previous research did find vulnerabilities in AMD CPUs (e.g., TakeAway [14] and Speculative Register Dereferencing [26]), it was commonly believed [18] that AMD processors are not vulnerable to Meltdown-type attacks.

In this paper, we challenge this belief and examine AMD processors for Meltdown-type behaviour. Originally, the authors of Meltdown [3] discussed that they could not trigger the kernel memory leak on AMD. However, they did not observe the exact reasons why the leak of kernel memory was not triggered. In the following white paper, [20] AMD reported that their microarchitecture load instructions will never fetch data if it is not architecturally allowed in the current execution context. Although it is true, we report another violation that is very similar to Meltdown-type behaviour. The violation we report does not lead to cross address space leaks, but it provides a reliable way to force an illegal dataflow between microarchitectural elements. Unlike the previous AMD vulnerabilities [14, 26], the flaw we report is the first flaw that proves that it is possible to force an illegal data flow between microarchitectural elements. The consequence of having a code snippet vulnerable to such behaviour may allow an attacker to poison the transient execution of the AMD CPU from the microarchitectural element. In addition, this discovery shows that AMD does implement speculation on memory accesses similar to Meltdown-type attacks, suggesting that even more, similar flaws might be yet to unveil.

*Contributions.*

- A novel transient injection primitive for AMD processors and analysis of its limitations.
- Discussion of an application of the primitive to enhance other, Spectre-type vulnerabilities.

*Responsible Disclosure.* We shared our findings with the AMD Security team on October 29th 2020. AMD team assigned a CVE-2020-12965 with Medium severity and published a bulletin with mitigation recommendations [27].

## 2 BACKGROUND

Fig. 1 present a general overview of the AMD Zen2 microarchitecture. It consists of three stages: 1) In-order frontend, that fetches and decodes instructions; 2) Out-of-Order Execution Unit 3) The Load-Store Unit.

This section presents a background of cache attacks and transient execution attacks.

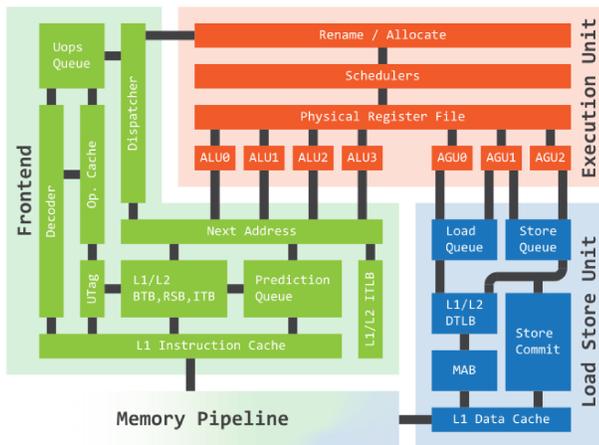

Figure 1: Zen2 microarchitecture.

### 2.1 Caches

Due to significant latency of memory access CPUs are equipped with the intermediate memory buffer to host recently accessed data to reduce the memory latency. These buffers are called caches. Usually, CPUs have multiple levels of caches, when the fastest level has the smallest size, and the slowest level has the largest size. Zen2 has a separate L1 cache for data(L1D) and instructions(L1i), L2 is a unified cache that is private per each core and L3 is also unified, while it is shared across all cores.

Since caches are supposed to decrease the latency, they inherently introduce the timing side-channel in systems. There were many cache attacks discovered over past years [7][10][17][2][5]. These attacks rely on two main technique "Flush+Reload" and "Prime+Probe". "Flush+Reload" measure access time of the element and check the target location of the element, based on the latency. The scope of the attack is limited, due to limited access to the shared elements between an attacker and a victim. This technique also allows us to check the current microarchitectural state changes.

An attacker may flush the element, and execute code, which depends on some secret, then the attacker can check the state of the element to detect whether it was cached. Based on the location of the element, the attacker may reduce the entropy of the secret. "Prime+Probe" targets shared cache (maybe L1 and LLC as well) and monitors which elements from cached sets were evicted during the unit of time. Such a strategy allows an attacker to assume which elements were cached by a victim.

### 2.2 Out-Of-Order Execution and Microarchitectural elements

Due to different and complex workload, dynamic scheduling or out-of-order execution is crucial to increase the instruction level-parallelism. Instructions get decoded and enter the processor pipeline in-order, then they are issued to the execution units out of order (this is also called to be executed transiently), then instructions leave the execution unit in-order and the result of each instruction becomes architecturally visible. Each instruction on the pipeline has its state (execution state, operands, result). To trace the state of each instruction, the common practice is that processors have many microarchitectural buffers to keep the intermediate state. In particular, x86 architecture has dedicated buffers for memory operations(Load, Store, cache misses, page table walk). Some of these buffers work completely transparently to the ISA (Store/Load Buffers, Special Register Buffers, etc.), some of them on the other hand are controlled by the ISA (Translation Lookaside Buffer, Memory Caches, etc).

*Load Queue.* To keep track of the Load operations issued by the CPU pipeline, a processor may have a dedicated buffer. Every time load instruction is issued, it populates the load Buffer. In AMD Zen-Family processors, such a Load buffer is called Load-Queue(Figure 1). As soon as the entry is populated in the Load buffer, it may be delivered the data from the cache or the memory.

*Store Queue.* As well as tracking load operations, tracking store operations is necessary. A dedicated buffer for this purpose is called Store Buffer(in Zen-family it is Store-Queue). In x86 SB entries are resolved strictly in-order to not break the sequential consistency property, unless it is a special write operation (such as non-temporal moves).

*TLB.* To keep independent address space per each process, x86 ISA keeps the translations between logical(virtual) addresses and physical addresses on the machine. These translations are encoded in the data structure called page-tables. To optimize the translation process CPUs are equipped with a TLB - Translation Lookaside Buffer, which keeps set-associative translation between virtual and physical addresses. TLB is partially managed by the explicit instructions in the ISA.

As it is shown on practice [9, 11, 13], there may be numbers of optimizations on dataflow between buffers, which work transparently.

### 2.3 Transient execution attacks

When instructions are executed out-of-order, some instructions may be executed, when they are not supposed to. This happens, when the processor speculates(branch predictions, return address prediction, etc.), delays the exception handling, needs microcode assistance. In such a case, all dependent instructions have to leave the pipeline to make the redispatch of the correct instructions stream with correct operands. Before instructions leave the pipeline, they still may be executed on the execution units, but the result of execution is still not architecturally visible. However, the microarchitectural state changes accordingly. This opens a covert channel since changes in the microarchitectural level are visible in different domains (such as timing).

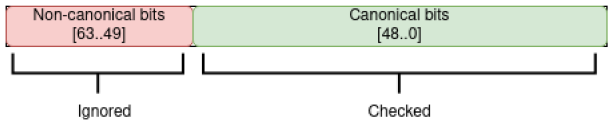

Figure 2: Partial address match

Many attacks [8, 9, 11, 13, 19] proved that it is possible to leak data from the architectural buffers and inject data as well by altering the microarchitectural state and analyzing it afterwards.

*Meltdown.* Meltdown attack exploits illegal dataflow in the transient execution window. The classical version of meltdown relies on Out-of-Order execution to bypass the general protection of the Intel CPU and access data from different address space. If data(of victim address space) is cached, it can be returned to the load instruction and be forwarded further(to following instructions) until abort happens (due to faulted load or any other reason).

Listing 1: Meltdown example.
```
// Attacker code
res = array1[*kern_addr*4096];
```

Listing 1 presents a code snippet, where the attacker just runs code in its userspace and accesses *kern_addr*, while the access will fault, the following instruction (dereference element of *array*1) will be executed. We multiply the content of the kernel address by 4096 to avoid cache line collision and L1D hardware prefetching. IF one runs such a code on a processor vulnerable to a Meltdown, it is possible to mount "Flush+Reload" afterwards in order to detect the locations of array1 elements and detect which value was stored in the kernel address.

*MDS.* While classical Meltdown attack targets fetching data from L1D cache, MDS attacks[12] class consists of many different attacks each of them targets a specific microarchitectural buffer. Under different circumstances, a faulted instruction may be forwarded data from Store-Buffer, Load-Buffer, Line-Fill Buffer, etc.

Both classical meltdown and MDS attacks are considered[23] to be Meltdown-type attacks.

So, *is Meltdown-type of attacks possible on AMD processors?* We give a negative answer to that question, however, we note that another similar to Meltdown attack is possible.

## 3 OVERVIEW

We report a flaw in AMD CPUs, which leads to illegal data propagation within the instruction stream, based on a partial address match. Our observation of transient loads corresponds to the behaviour reported by a white paper from AMD [20]: For a load instruction issued onto the pipeline to work a TLB hit is required. But we noticed, that for a transient load to work its enough to have only the canonical part of the address to be matched ([47..0] bits of the virtual address). This means that if we try to dereference the non-canonical VA1 when TLB contains an entry with canonical VA2 ([47..0] bits of VA1 and VA2 match) the content of the VA2 will be passed transiently to the load (Figure 2).

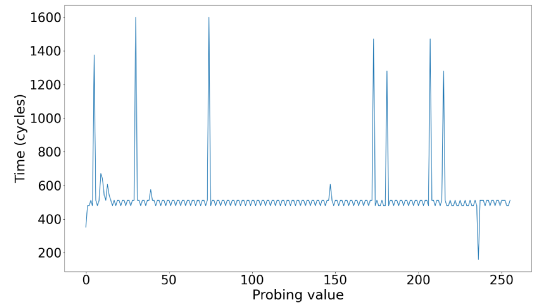

Figure 3: Signal

We suspect that the full address check is done when instruction leaves the pipeline in program order. Listing 2 provides the basic code snippet, which triggers the violation in the transient execution path. The result of the execution is measured with "FLush+Reload" primitive (Figure 3).

Listing 2: Non-canonical address violation.
```
// addr2 is non-canonical address;
addr2 = addr1 | 0xff0f000000000000;
addr1[offset] = SECRET_VAL;

// SECRET_VAL is leaked here
oraclearr[4096 * addr2[OFFSET]];
```

Hereafter, we refer to this finding as a non-canonical address violation. We verified, that source of leakage could be the L1 cache and a not-committed entry from the Store Queue as well. AMD Optimization manual [16] describes, that [11:0] bits are used to determine the Store-To-Load-Forwarding (STLF). However, we did not see any illegal STLF which is triggered by the lowest 12-bit match even within the same address space. Moreover, to trigger the illegal STLF we noticed, that TLB-hit is not enough. The second condition is store instruction from the Store Queue has to be an L1 DCache hit (Figure 4).

We verified our main observation (non-canonical address violation) on both speculative (Spectre-type) and non-speculative (Meltdown-type) execution paths. To explain this behaviour we learned the patent[25] where it is stated, that AMD CPUs may require the micro-TLB hit before any load instruction passes Figure 4.

micro-TLB is a dedicated structure, which keeps partial information from the main TLB. However, we were told by the AMD security team, that the Ryzen-family of CPU is not equipped with micro-TLB, but use the main TLB for this check. In other words, if there is no TLB hit, Load will not be passed transiently. Hence, we found out that the main TLB ignores the upper bits when it compares.

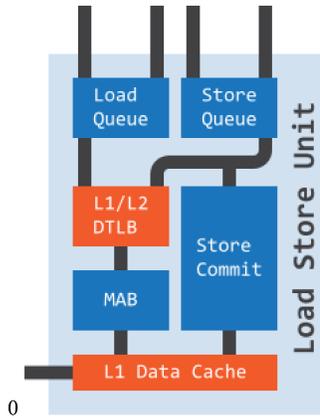

**Figure 4: Datapath checks (a μarch element is red if check is performed)**

**Listing 3: Vulnerability example.**
```c
int do_something_with_ptr(void* addr){
    if(addr < BASE_ADDR) return -EINVAL;

    // ...addr is used somewhere here
    return 0;
}
```

We also could not trigger this leak between two user spaces running on the same core, obviously, because of the TLB flush. However, it is possible to "leak" across different threads with the same address space via L1D Cache.

## 4 EXPLOITATION

While the basic primitive itself is useless to leak data from the victims address space, it provides the ability to make a victim's code to inject wrong data internally. The basic idea of the violation is explained by the code snippet in Listing 3. As you see, if address *addr* passes the check to be larger or equal to *BASE_ADDR*, because of the non-canonical violation it still can transiently pass a value from the address smaller than *BASE_ADDR*.

### 4.1 Sandbox

Since this primitive does not work in cross-address space (neither user-user nor user-kernel) scenarios, we focus our attention on sandbox scenario and injections here. We enumerate conditions, which have to be kept to make this mechanism exploitable:

- First, code has to not instrument the highest (non-canonical part) of virtual addresses.
- Second, direct access to some memory should not be programmatically allowed due to a control check (normal address should be mapped, otherwise TLB hit will not occur).

We studied different software (Spidermonkey, latest Linux Kernel) and we could not find gadgets to exploit it. Therefore, we give a possible code snippet, which could be exploited to trigger the illegal dataflow in the sandbox.

Lets look at the synthetic scenario of the sandbox code gadget:

**Listing 4: Vulnerability example.**
```c
int dereference_addr(void* addr){
    if(addr < SANDBOX_BASE_ADDR)
        return -EINVAL;
    access_addr(addr);
    return 0;
}
```

In this example, we may have *lfence* to prevent speculative execution of the *access_addr* function, but the violation comes from the fact, that even if address *addr* passed the check, it does not mean, that its value comes from the higher addresses indeed. If an attacker can "overflow" (though real overflow of the pointer doesn't happen) the value to make the address to be non-canonical, the attacker can still inject values from the lower addresses (lower than SANDBOX_BASE_ADDR) and transiently inject the attacker-controlled value into the victim sandbox domain. Such an injection can amplify other side-channel attacks based on classical Spectre. We argue that it's not easy to catch such a violation because there is no widely adopted software runtime check for addresses to be canonical.

## 5 EVALUATION

We explored the AMD processor of the Zen-family. CPU models are listed in Table 1. We built a basic covert channel on top of given primitives and achieved the bandwidth of 125 bytes/s. Main of the delay for the covert channel comes from the synchronization and FLUSH+RELOAD phases.

**Table 1: AMD CPUs we tested**

| Model | Year | uarch | violation |
| --- | --- | --- | --- |
| EPYC7262 | 2019 | Zen2 | Yes |
| Ryzen_7_2700X | 2018 | Zen+ | Yes |
| Threadripper2990WX | 2018 | Zen+ | Yes |

We also tested Intel CPUs for such behaviour. All Intel CPUs that are vulnerable to MDS attacks inherently have the same flaw described here. We tested one MDS-resistant *Intel(R) Core(TM) i7-10510U*, and we did not detect such a flaw.

## 6 CONCLUSIONS AND DISCUSSION

We present the flaw in the AMD processors based on non-canonical address violation. We analyze the limitation of some transient memory accesses. We verified the hardware flaw is presented on Zen+, Zen2 families of AMD CPUs. We also discuss potential scenarios when the flaw may become a vulnerability, how it can be triggered and how it can enhance other side-channel attacks.